\documentstyle[12pt,a4,cite,epsfig]{article}

\newcommand{\ba}{\begin{eqnarray}}
\newcommand{\ea}{\end{eqnarray}}
\newcommand{\be}{\begin{equation}}
\newcommand{\ee}{\end{equation}}
\newcommand{\dis}{\displaystyle}

\begin{document}

\begin{titlepage}
\begin{flushright}
CAFPE-119/09\\
UG-FT-249/09\\
\end{flushright}
\vspace{2cm}
\begin{center}

{\large\bf Hadronic Light-by-Light Contribution to Muon g-2
\footnote{Contributed to ``KLOE2 Physics Workshop'', April 9-10  2007, Frascati, Italy.}}\\
\vfill
{\bf  Joaquim Prades}\\[0.5cm]

CAFPE and Departamento de
 F\'{\i}sica Te\'orica y del Cosmos, \\
Universidad de Granada, 
Campus de Fuente Nueva, E-18002 Granada, Spain.\\[0.5cm]

\end{center}
\vfill

\begin{abstract}
\noindent
I present the main results obtained in a recent work 
together with Eduardo de Rafael and Arkady Vainshtein
on the hadronic light-by-light contribution  to 
 muon g-2. We came to the estimate
 $a^{\rm HLbL}_\mu=(10.5\pm2.6) \times 10^{-10}$. 
 Here, some emphasis is put in pointing out where the future
KLOE2 two-photon experimental program can help to reduce 
the present model dependence of $a^{\rm HLbL}_\mu$.
\end{abstract}
\vfill
May 2009
\end{titlepage}
\setcounter{page}{1}
\setcounter{footnote}{0}

\section{Introduction}
\label{intro}
\begin{figure}[hbt]
\begin{center}
\epsfig{file=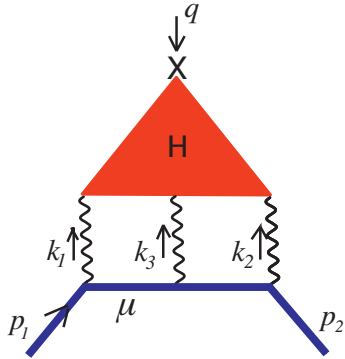,width=4.5cm}
\end{center}
\caption{Hadronic light-by-light scattering contribution.}
\label{fig:1}      
\end{figure}

One of  the six possible photon momenta configuration to the 
hadronic light-by-light  (HLbL) contribution  to the muon anomalous 
 magnetic moment  $a=(g_\mu-2)/2$ is shown in Fig. \ref{fig:1} 
and described by the  vertex function 
\ba
\label{Mlbl}
\dis{\Gamma^\mu} (p_2,p_1)
&=&  - e^6  
\int {{\rm d}^4 k_1 \over (2\pi )^4}
\int {{\rm d}^4 k_2\over (2\pi )^4}  
{\Pi^{\mu\nu\rho\sigma} (q,k_1,k_3,k_2) 
\over k_1^2\, k_2^2 \, k_3^2} \nonumber \\ &\times&  
\gamma_\nu (\not{\! p}_2+\not{\! k}_2-m )^{-1} 
\gamma_\rho (\not{\! p}_1-\not{\! k}_1-m )^{-1} \gamma_\sigma \, 
 \nonumber \\ 
\ea
where $q \to 0 $ is the momentum of the
photon that couples to the external magnetic source,
$q=p_2-p_1=-k_1-k_2-k_3$ and $m$ is the muon mass. 

 The dominant contribution to the hadronic four-point  function 
\ba
\label{four}
\Pi^{\rho\nu\alpha\beta}(q,k_1,k_2,k_3)&=& \nonumber \\
i^3 \int {\rm d}^4 x \int {\rm d}^4 y
\int {\rm d}^4 z \, {\rm e}^{i (-k_1 \cdot x + k_2 \cdot y + k_3 \cdot z)} \, 
 \langle 0 | T \left[  V^\mu(0) V^\nu(x) V^\rho(y) V^\sigma(z)
\right] |0\rangle && 
\ea
comes from the three light quark 
$(q = u,d,s)$ components in the electromagnetic current
$V^\mu(x)=\left[ \overline q \widehat Q \gamma^\mu q \right](x)$
where $\widehat Q$ denotes the quark electric charge matrix. 
We are interested in the limit $q \to 0$ where
current conservation implies
\ba
\dis{\Gamma^\mu} (p_2,p_1) =
- \frac{a^{\rm HLbL}}{4 m} \, 
\left[\gamma^\mu, \gamma^\nu \right] \, q_\nu \, .
\ea

Here I would like to describe the main results
of \cite{PRV09}.  For recent reviews of previous work
\cite{HKS95,HK98,BPP95,BPP02,KN02a,KN02b,MV04}, 
see \cite{PR08,EdR09,JN09}.

\section{Numerical Conclusions and Prospects}
\label{sec:1}

The  discussion in \cite{PRV09} 
lead the authors to give 
the following numerical conclusions according to an $1/N_c$
expansion \cite{EdR94} --such expansion works reasonably well:

\begin{itemize}

\item \emph{Contribution from $\pi^0, \eta$ and $\eta'$ exchanges}:
Implementing a new OPE constraint
 into a neutral pion exchange model \cite{MV04}, the authors  of Ref. \cite{MV04} obtained  $(11.4 \pm 1.0) \times 10^{-10}$
for this contribution.  Within the  ENJL model
the momenta higher than a certain cutoff is accounted separately
via quark loops  \cite{BPP95,BPP02} while in the OPE based model 
these momenta are already included
into the result. Assuming that the bulk
 of high energy quark loops  are associated with pseudo-scalar 
exchange Ref. \cite{BPP95,BPP02} obtains
 $(10.7 \pm 1.3) \times 10^{-10}$ 
after adding these to the neutral pion
exchange within the ENJL model. Taking into account 
this discussion, the authors of \cite{PRV09} quote as 
central value the one in \cite{MV04}  with the largest error quoted 
in \cite{BPP95,BPP02}:
\be
a^{\rm HLbL}(\pi^0,\eta,\eta')=
  (11.4 \pm 1.3) \times 10^{-10} \, .
\ee

\item \emph{Contribution from pseudo-vector exchanges}:
The analysis done in  \cite{PRV09} 
suggests that the errors quoted within 
the large $N_c$ ENJL model are underestimated. 
Taking the average within both
estimates and  raising the present uncertainty to cover both, Ref.\cite{PRV09} quote
\be
a^{\rm HLbL}({\rm pseudo-vectors})
 = (1.5 \pm 1.0 ) \times 10^{-10} \, .
\ee

\item \emph{Contribution from scalar exchanges}:
The ENJL model should give a good estimate 
of these large $N_c$ contributions, 
the authors of \cite{PRV09} therefore keep the result from
 \cite{BPP95}  but with a larger conservative
error to cover for other unaccounted higher resonances that give  negative  contributions:
\be
a^{\rm HLbL}({\rm scalars}) = -(0.7 \pm 0.7 ) 
\times 10^{-10} \, .
\ee

\item \emph{Contribution from dressed  pion and kaon loops}:
 The next-to-leading in $1/N_c$ contributions are 
the most complicated to calculate at present. 
In particular, the charged pion loop shows
a large instability due to model dependence. 
This and the contribution of higher resonances loops 
was taken into account  in \cite{PRV09} by 
taking the central value as the full VMD result quoted \cite{BPP95} 
with again a large conservative error:
\be
a^{\rm HLbL}(\pi^+-{\rm dressed \,\, loop}) 
= -(1.9 \pm 1.9 ) \times 10^{-10} \, .
\ee

\end{itemize}

Adding the contributions above and   errors in quadrature, 
as well as the small charm quark contribution \\
$a^{\rm HLbL}({\rm charm})= 
0.23 \times 10^{-10}$,
  one gets our best estimate \cite{PRV09}
\be
a^{\rm HLbL} = (10.5 \pm 2.6) \times 10^{-10} \, .
\ee

The proposed new $g_\mu-2$ experiment 
accuracy goal of $1.4 \times 10^{-10}$ calls
for a considerable improvement in the present
calculations. The use of further theoretical and experimental constraints could result in reaching such accuracy soon enough. In particular, imposing as many as possible short-distance QCD constraints
\cite{HKS95,HK98,BPP95,BPP02,KN02a,MV04} has result in a better understanding of the numerically dominant $\pi^0$ exchange. 
At present, none of the  light-by-light hadronic parametrization satisfy fully all short distance QCD constraints. In particular, this requires the  inclusion of
 infinite number of narrow states for other than two-point functions  and two-point functions with soft insertions \cite{BE03}.
A numerical dominance of certain momenta configuration
can help to minimize  the effects of short distance QCD constraints
 not satisfied, as in the model in \cite{MV04}.

Recently, an off-shell form factor for the $\pi^0$ neutral
exchange has been discussed in \cite{NY09} to get
$a^{\rm HLbL}_\mu$ -- the numerical values for the $\pi^0$ exchange obtained 
are very similar to the ones quoted above. 
How to take off-shellness effects consistently in the full 
four-point function (\ref{four}) remains however an open question 
\cite{NY09}.

More experimental information on the decays $\pi^0 \to \gamma \gamma^*$,
$\pi^0 \to \gamma^* \gamma^*$ and $\pi^0 \to e^+ e^-$ 
(with radiative 
corrections included \cite{RW02,KKN06,KM09}) 
can also help to confirm some of the neutral pion exchange results.

A better understanding of other smaller 
contributions but with comparable uncertainties needs both more
 theoretical work and  experimental information.
This refers in particular to  pseudo-vector exchanges. 
Experimental data on radiative decays and two-photon 
production of these and  other C-even resonances can be useful in that respect. 
Experimental information on processes $\pi^0\pi^0 \to \gamma^*\gamma^*$  
and $\pi^+\pi^- \to \gamma^*\gamma^*$   would be very welcome for that.
 For instance, these processes
are  related to the two-photon coupling of the lightest QCD resonance
--the $\sigma$ \cite{PE07,OR08,BP08,MN08}.

New approaches to the pion dressed loop contribution, together with experimental
information on the vertex $\pi^+\pi^-\gamma^*\gamma^*$  would 
also be very welcome.
Measurements of two-photon processes like 
$e^+e^- \to e^+e^-\pi^+\pi^-$ can be useful to give information on that 
vertex and again could reduce the model dependence.

 The two-gamma physics program at KLOE2 will be very useful
 and well suited  in the processes mentioned above which
 information can help 
to decrease the present model dependence of $a^{\rm HLbL}_\mu$.

\section*{Acknowledgements}

 It is a pleasure  to thank very enjoyable 
collaborations with Hans Bijnens, Elisabetta Pallante, Eduardo de Rafael and Arkady Vainshtein. 
Work supported in part by MICINN, Spain and FEDER, European Commission
(EC) Grant No. FPA2006-05294, by the Spanish Consolider-Ingenio 2010 Programme
CPAN Grant No. CSD2007-00042, by Junta de Andaluc\'{\i}a Grants
 No. P05-FQM 347 and P07-FQM 03048 and by the EC RTN FLAVIAnet Contract
No. MRTN-CT-2006-035482.
                     

%


\begin{thebibliography}{}
%
%

\bibitem{PRV09}
J. Prades, E. de Rafael and A. Vainshtein in 
\textit{Lepton Dipole Moments}, B.L. Roberts and W.J. Marciano, (eds)
 (World Scientific, Singapore, 2009) 309-324,
arXiv:0901.0306.

\bibitem{HKS95}
M. Hayakawa, T. Kinoshita and  A.I. Sanda, 
Phys. Rev. Lett. \textbf{75}
(1995) 790; 
Phys. Rev. D \textbf{54} (1996) 3137.

\bibitem{HK98} M. Hayakawa and T. Kinoshita, 
Phys. Rev. D \textbf{57} (1998)465;
 Erratum-ibid. \textbf{66} (2002) 073034. 

\bibitem{BPP95}
J. Bijnens, E. Pallante and J. Prades, 
Nucl. Phys. B \textbf{474} (1996) 379;
Phys. Rev. Lett. \textbf{75} (1995) 1447; 
Erratum-ibid. \textbf{75} (1995) 3781.

\bibitem{BPP02}
J. Bijnens, E. Pallante and J. Prades, 
Nucl. Phys. B \textbf{626} (2002) 410.

\bibitem{KN02a}
M. Knecht and A. Nyffeler, 
Phys. Rev. D \textbf{65} (2002) 073034.

\bibitem{KN02b}
M. Knecht, A. Nyffeler, M. Perrottet and E. de Rafael,ç
 Phys. Rev. Lett. \textbf{88} (2002) 071802.

\bibitem{MV04}
K. Melnikov and A. Vainshtein,
Phys. Rev. D \textbf{70} (2004) 113006.

\bibitem{PR08}
J. Prades,  Nucl. Phys. B (Proc.\ Suppl.)  
\textbf{181-182} (2008) 15;
J. Bijnens and J. Prades, Mod. Phys. Lett. A 
\textbf{22} (2007) 767;
  Acta Phys.\ Polon.\  B \textbf{38} (2007) 2819.

\bibitem{EdR09} E. de Rafael,
  Nucl.\ Phys. B  (Proc.\ Suppl.)  \textbf{186} (2009) 211;
  D.W. Hertzog {\it et al.},
  arXiv:0705.4617;
J.P. Miller, E. de Rafael and B.L. Roberts,
  Rept.\ Prog.\ Phys.\  {\bf 70} (2007) 795.

\bibitem{JN09}
  F. Jegerlehner and A. Nyffeler,
  arXiv:0902.3360;
F. Jegerlehner,  Lect.\ Notes Phys.\  \textbf{745} (2008) 9;
  Acta Phys.\ Polon.\  B {\bf 38} (2007) 3021.

\bibitem{EdR94}
E. de Rafael, Phys. Lett. B \textbf{322} (1994) 239.

\bibitem{BE03}
J. Bijnens {\it et al.}
JHEP \textbf{04} (2003) 055.

\bibitem{NY09}
A. Nyffeler,   arXiv:0901.1172.

\bibitem{RW02}
M. Ramsey-Musolf and M.B. Wise, Phys. Rev. Lett.
\textbf{89} (2002) 041601.

\bibitem{KKN06}
K. Kampf, M. Knecht and J. Novotny,
  Eur.\ Phys.\ J.\  C \textbf{46} (2006) 191.

\bibitem{KM09}
  K. Kampf and B. Moussallam,
  Phys. Rev. D \textbf{79} (2009)  076005;
  K. Kampf,   arXiv:0905.0585.


\bibitem{PE07}
M.R. Pennington, Phys. Rev. Lett. \textbf{97} (2007) 011601.

\bibitem{OR08}
 J.A. Oller and L. Roca,
  Eur.\ Phys.\ J.\  A \textbf{37} (2008) 15;
J.A. Oller, L. Roca and C. Schat,
  Phys.\ Lett.\  B \textbf{659} (2008) 201.

\bibitem{BP08}
J. Bernab\'eu and J. Prades, Phys. Rev. Lett. 
\textbf{100} (2008) 241804.

\bibitem{MN08} 
G. Mennessier, S. Narison and W. Ochs,
Phys. Lett. B \textbf{665} (2008) 205.

\end{thebibliography}
\end{document}